%% file: FukayaLat09.tex
\documentclass{PoS}

\title{Exploring chiral dynamics with overlap fermions}

\ShortTitle{Exploring chiral dynamics with overlap fermions}

\author{\speaker{Hidenori Fukaya} for JLQCD \& TWQCD collaborations
\\
        Department of Physics, Nagoya University, 464-8602, Nagoya, Japan\\
        E-mail: \email{hfukaya@eken.phys.nagoya-u.ac.jp}}

\abstract{
This talk presents a lattice study of the spontaneous chiral symmetry
breaking performed by the JLQCD and TWQCD collaborations with
dynamical overlap fermions.  
Our lattice configurations are generated 
in a fixed topological sector. 
Since finite volume effects, partly due to the fixed global
topology, are mainly induced by pion fields,
the dependence on the lattice volume, topological charge
and quark masses can be analytically predicted using
chiral perturbation theory (ChPT).
We find a good agreement of Dirac operator spectrum calculated on the
lattice with the ChPT prediction including its finite size scalings,
through which the chiral condensate is determined with good accuracy.
}

\FullConference{The XXVII International Symposium on Lattice Field Theory - LAT2009\\
		 July 26-31 2009\\
		 Peking University, Beijing, China}

\begin{document}

\input{sec1.tex}

\input{sec2.tex}

\input{sec3.tex}
\input{sec4.tex}

\input{sec5.tex}

\input{ref.tex}
\end{document}

%% file: sec1.tex
\section{Introduction}
\label{sec:intro}

The spontaneous chiral symmetry breaking plays a central role
in the low-energy Quantum Chromodynamics (QCD).
It is understood that this phenomena is the source of the hadron masses
of order $\Lambda_{\mathrm{QCD}}$, the QCD scale.
An important exception is the pion, which is nearly massless, as it is
a pseudo-Nambu-Goldstone boson.
The pion dynamics is well described by an effective theory, known
as chiral perturbation theory (ChPT) \cite{Gasser:1983yg}, which is
constructed based on the pattern of spontaneous symmetry breaking.

Parameters in ChPT are not known a priori.
In phenomenological analysis, they are determined with experimental
data as inputs, but it is more desirable if they can be calculated
starting from the first-principles of QCD. 
This sets a challenge for lattice QCD.
At the leading order, there are two parameters: the chiral condensate
$\Sigma$ and pion decay constant $F$ in the chiral limit.
Calculation of these parameters has long been one of the main
issues in lattice QCD.
In particular, the calculation of $\Sigma$ has been notoriously
difficult, as it survives only in the thermodynamical limit,
{\it i.e.} the limit of massless sea quarks after taking infinite
volume limit.
The determination of other parameters in ChPT, such as the low energy
constant at the next-to-leading (NLO) order can be done only after the
leading order parameters are determined precisely.

Lattice QCD has become the most powerful tool for non-perturbative
calculation of strong interaction of hadrons, with the help of the
rapid speed-up of computers.
In fact, lattice QCD has even played a leading role in the development
of high-end computers.
Still, the mechanism of chiral symmetry breaking remained not entirely
clear until recently, since the chiral symmetry itself was violated in
the simulations with the conventional lattice fermion formulations.
It is theoretically known that the use of Neuberger's overlap-Dirac
operator \cite{Neuberger:1997fp} is a solution to this problem 
as it realizes exact chiral symmetry at finite lattice spacings
\cite{Ginsparg:1981bj,Luscher:1998pq}.
Because of its numerical cost, however, it was only recently
that the large-scale simulation of dynamical overlap fermions became
feasible.

The numerical cost of the overlap-Dirac operator is high, compared to
other non-chiral or non-flavor-symmetric lattice fermions, as it
involves an approximation of the sign function of the hermitian
Wilson-Dirac operator.
The cost increases even more
when the Atiyah-Singer index of the Dirac operator, which corresponds
to the topological charge of the background gauge field, changes its
value by $\pm$1.
This is because the molecular dynamics steps have to involve an extra
procedure \cite{Fodor:2003bh} in order to catch a sudden jump of the
fermion determinant on the topology boundary.
This additional procedure, known as the reflection/refraction, needs
numerical cost potentially proportional to the lattice volume squared.

Recently, the JLQCD and TWQCD collaborations have performed
large-scale simulations of 2- and 2+1-flavor QCD
employing the overlap fermions for sea quarks \cite{Aoki:2008tq}.
We avoid the extra numerical cost due to the change of topology
by a modification of the lattice action to suppress the topology
tunneling as proposed in 
\cite{Izubuchi:2002pq, Vranas:2006zk, Fukaya:2006vs}.
Our lattice simulations are confined in a fixed topological
sector, so that an expectation value of any operator could be
deviated from the value in the true QCD vacuum.
This effect can be understood as a finite volume effect and estimated 
in a theoretically clean manner as discussed below.
It is worth noticed that the simulation parameters contain those in
the $\epsilon$-regime on a $L\sim 2$~fm lattice, as well as in the
conventional $p$-regime
\cite{Fukaya:2007fb, Fukaya:2007yv, Fukaya:2007pn}.
This enables us to study the chiral dynamics in an entirely different
set-up and to determine the low-energy constants at the point very
close to the chiral limit.

With exact chiral symmetry, the study of spontaneous chiral symmetry
breaking is theoretically clean, but it still requires a good control
of the systematic effects due to the finite volume \cite{DeGrand:2006nv}.
For such infra-red effects, the lightest particle, which is the pion,
gives a dominant contribution.
It should therefore be possible to use analytic calculations within
ChPT in order to predict the finite volume corrections for a quantity
of interest. 
Then, the lattice results can be directly fitted with these finite-volume
formulae of ChPT to determine the relevant low-energy constants.
The effect of fixed topology can also be understood as one of such
infra-red effects since the {\it global} topological charge should not
affect the physics at a local sub-volume when the entire volume $V$ is
large enough \cite{Brower:2003yx,Aoki:2007ka}. 
In a calculation of the topological susceptibility
\cite{Aoki:2007pw,Chiu:2008kt, Chiu:2008jq, Hsieh}
through topological charge density correlator,
we can actually see that local topological excitations are active even
when the global topological charge is kept fixed.
Its result is consistent with an expectation of ChPT, which implies
that the ChPT-based analysis is valid for the effects due to the fixed
topological charge\cite{Mao:2009sy, Aoki:2009mx}.

There have been a number of analytical works that aimed at controlling
the infrared effects occurring in the lattice simulations.
A well-known example is the finite volume correction due the pions
wrapping around the lattice \cite{Bernard:2001yj,Colangelo:2005gd}.
Extended works are necessary when the system enters the so-called 
$\epsilon$-regime \cite{Gasser:1987ah,Leutwyler:1992yt} by reducing
the sea quark mass to the vicinity of the chiral limit.
In this regime, the vacuum fluctuation of the pion field plays a
special role and a non-perturbative approach is needed in ChPT.
Namely, the zero-momentum pion mode has to be integrated over the
group manifold of the chiral symmetry in contrast to the case of the
conventional $p$-regime where a certain vacuum is (randomly) chosen by
the spontaneous symmetry breaking.
Recently, the partition functions with fully non-degenerate flavors
\cite{Splittorff:2002eb} were calculated, so that even 
the (partially) quenched analysis \cite{Bernardoni:2007hi} of the 
meson correlators is possible.
To study more realistic set-up, {\it i.e.} including the strange quark
in the $p$-regime, several hybrid method to treat both the $\epsilon$-
and $p$-regimes have been proposed
\cite{Bernardoni:2008ei,Damgaard:2008zs}. 
The effect of fixed topology is worked out in 
\cite{
Aoki:2009mx}.
We also note that the effects of explicit violation of chiral symmetry
due to the Wilson term are also discussed \cite{Bar:2008th,Shindler:2009ri}, 
which is needed to study the Wilson fermion simulations near the chiral limit
\cite{Hasenfratz:2008ce, Giusti:2008vb}.

In this talk, the dynamical overlap fermion simulation 
by the JLQCD and TWQCD collaborations
is reviewed in Section~\ref{sec:ov}.
In Section~\ref{sec:ChPT}, we discuss the finite size scaling
as well as the global topological effects within ChPT.
As an example, our recent result for chiral condensate 
\cite{:2009fh, Hashimoto:2009iv, Hashimoto:2009iy}
is presented in Section \ref{sec:cond}.
Summary and conclusion are given in Section \ref{sec:summary}.

%% file: sec2.tex
\section{Dynamical overlap fermion at fixed topology}
\label{sec:ov}

We employ the overlap-Dirac operator 
\cite{Neuberger:1997fp} 
\begin{equation}
  \label{eq:ov}
  D(m) = 
  \left(m_0+\frac{m}{2}\right)+
  \left(m_0-\frac{m}{2}\right)
  \gamma_5 \mbox{sgn}[H_W(-m_0)],
\end{equation}
for the quark action. Here $m$ denotes the quark mass and
$H_W\equiv\gamma_5D_W(-m_0)$ is the Hermitian
Wilson-Dirac operator with a large negative mass $-m_0$.
We take $m_0=1.6$ throughout our simulations.
(Here and in the following the parameters are 
given in the lattice unit.)
In the chiral limit $m\to 0$, 
the overlap-Dirac operator (\ref{eq:ov}) satisfies 
the Ginsparg-Wilson relation \cite{Ginsparg:1981bj}
\begin{equation}
  \label{eq:GW}
  D(0)\gamma_5 + \gamma_5D(0)=\frac{1}{m_0}D(0)\gamma_5D(0).
\end{equation}
With this relation, the fermion action constructed from
(\ref{eq:ov}) has exact chiral symmetry under a
modified chiral transformation \cite{Luscher:1998pq}.
Moreover, it is known that the overlap-Dirac operator has
an index which corresponds to the topological charge
in the continuum limit \cite{Hasenfratz:1998ri}.

In the numerical implementation of the overlap-Dirac operator
(\ref{eq:ov}), the profile of near-zero modes of the kernel 
operator $H_W(-m_0)$ 
largely affects the numerical cost of the overlap fermion
(The presence of such near-zero modes is also a problem for
the locality property of the overlap operator \cite{Hernandez:1998et}.).
For the approximation of the sign function in 
(\ref{eq:ov}), the number of operations of the Wilson-Dirac operator
needed to keep a certain precision monotonically increases as the
condition number 
$\lambda^{max}_{W}/\lambda^{min}_{W}$ grows,
where $\lambda^{max/min}_W$ denotes 
the maximum/minimum eigenvalue of the operator $|H_W(-m_0)|$.
Moreover, since the overlap-Dirac operator is not uniquely determined
when $H_W(-m_0)$ has a zero eigenvalue, 
the overlap fermion determinant has a discontinuity.
This discontinuity of the determinant prevents 
smooth evolution of the molecular dynamics steps and
requires a special treatment, 
known as the reflection/refraction procedure \cite{Fodor:2003bh}.
It needs an extra numerical cost, which is potentially 
proportional to the lattice volume squared.

At currently available lattice spacings with 
conventional gauge actions,
the spectral density $\rho_W(\lambda_W)$ of the
operator $H_W(-m_0)$ is
non-zero at zero eigenvalue $\lambda_W$ = 0 \cite{Edwards:1998sh}.
Note that the appearance of $\rho_W(\lambda_W=0)$ is, however,
a lattice artifact due to the so-called dislocations: 
local lumps of gauge configurations \cite{Berruto:2000fx},
which disappears in the continuum limit.

To avoid the problem of the large extra numerical cost and of the
potentially ill-defined overlap operator,
we introduce additional Wilson fermions and
twisted-mass bosonic spinors to generate a weight
\begin{equation}
  \label{eq:detHw}
  \frac{\det[H_W(-m_0)^2]}{\det[H_W(-m_0)^2+\mu^2]},
\end{equation}
in the functional integrals 
\cite{Izubuchi:2002pq, Vranas:2006zk, Fukaya:2006vs}.
Both of fermions and ghosts are unphysical 
as their masses are of order of the lattice
cutoff, and thus do not affect low-energy physics.
The numerator suppresses the appearance of near-zero modes, 
while the denominator cancels unwanted effects from higher modes.
The ``twisted-mass'' parameter $\mu$ controls the value below which
the eigenmodes are suppressed. 
In our numerical studies, we set $\mu$ = 0.2.

\begin{figure}[tbp]
  \centering
  \includegraphics[width=10cm,clip=true]{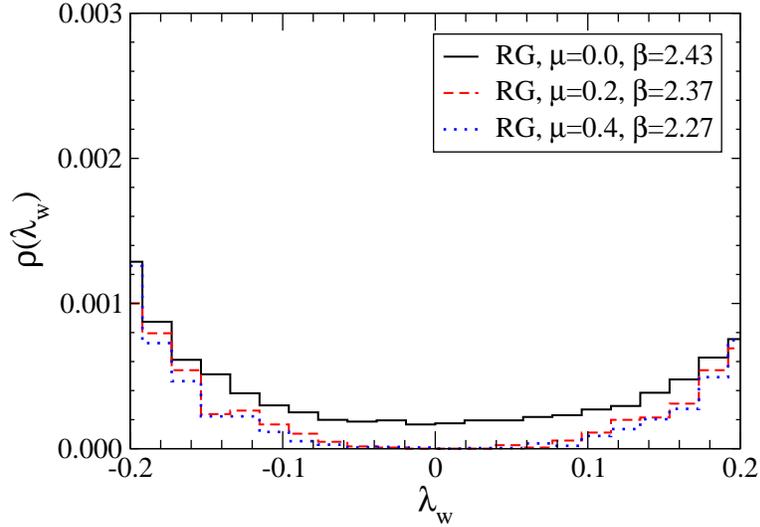}
  \caption{
    Histogram of the spectral density of $H_W(-m_0)$.
    Data for three values of $\mu$ ($\mu$ = 0.0, 0.2, and
    0.4) are shown in the plot.
    Note that $\mu=0$ corresponds to the case where the extra
    fermion determinant is turned off.
    }
  \label{fig:eigen_hist}
\end{figure}

As Fig.~\ref{fig:eigen_hist} shows, 
the near-zero modes of $H_W(-m_0)$
are actually washed out when $\mu$ is non-zero 
in quenched QCD simulations.
This leads to a large reduction of the numerical cost 
to approximate the sign function in (\ref{eq:ov}) \cite{Fukaya:2006vs}. 
We also find that the molecular dynamics evolution
is smooth in the hybrid Monte Carlo updates
and we can turn off the
reflection/refraction procedure.

The presence of zero-mode of $H_W(-m_0)$ is related to a topology
change:
the Atiyah-Singer index or the topological charge of gauge fields
changes its value when an eigenvalue $H_W(-m_0)$ crosses zero.
The condition $H_W(-m_0)=0$, thus, forms a topology boundary
on the gauge configuration space.
With the lattice action including (\ref{eq:detHw}), therefore, 
the topological charge never changes
during the molecular dynamics steps of the Hybrid Monte Carlo
(HMC) simulations.
In this work, the simulations are mainly performed in the trivial
topological sector $Q=0$.
In order to check the topological charge dependence, 
we also carry out independent simulations at $Q=+1$, $-2$ and $-4$
at some parameter choices.
The configuration space of a given fixed topology is simply 
connected in the continuum limit, hence it is natural to assume 
that the ergodicity of the Monte Carlo simulation is satisfied within 
in a given topological sector. 

In the Monte Carlo simulations, we choose 5-6 
different points of the up and down quark mass $m_{ud}$
in a range $0.002 \leq m_{ud}\leq 0.100$.
For the $N_f=2+1$ runs,  two values of the strange quark mass:
$m_s=0.080$ and 0.100 are taken.
For the gauge part, we use the Iwasaki gauge action \cite{Iwasaki:1985we}
at $\beta=2.3$ (except for the case of
$m_{ud}=0.002$ in the $N_f=2$ run where $\beta=2.35$ is chosen).
The lattice volumes are $V=L^3T=16^3\times 32\;(N_f=2)$
and $V=L^3T=16^3\times 48\;(N_f=2+1)$.
For the latter, we also carry out a run on a 
$V=L^3T=24^3\times 48$ lattice at $m_{ud}=0.025$ and $m_s=0.080$,
in order to check the finite volume effect.
The lattice scales $a^{-1}=1.667$ GeV ($N_f=2$) and 
$a^{-1}=1.833$ GeV ($N_f=2+1$) are determined 
from the heavy quark potential, using $r_0=0.49$ fm as an input
\cite{Sommer:1993ce}.
The lattice size is then estimated as $L\sim 1.9$ fm for
$N_f=2$, and  $L\sim 1.7$ fm for $N_f=2+1$ runs.
Note that for the lightest quark mass $m_{ud}=0.002$ $\sim3$ MeV, 
the system of pions is inside the $\epsilon$-regime. 

Since our gauge configurations are generated
in a fixed topological sector,
expectation value of any operator
could be different from those in the QCD vacuum.
Also, our lattice size is $\sim 2$ fm and
considerable finite volume effects, especially in the 
$\epsilon$-regime, are expected.
As our lattice size is, however, still kept 
larger than the inverse of QCD scale,
{\it i.e.} $\Lambda_{QCD}L \gg 1$,
both effects can be considered as a part of
infra-red physics for which pions 
are most responsible.
We, therefore, expect that chiral perturbation theory
(ChPT) can correct these systematic effects.
In the next section, we discuss how to evaluate 
physical observables in a fixed topological sector within ChPT
at finite $V$.
Non-perturbative treatment of the zero momentum mode, 
as well as the Fourier transform 
with respect to the vacuum angle $\theta$, play a key role.
Using their analytic formulae, we can 
convert the lattice QCD results on a finite lattice to 
the values in the true QCD vacuum 
in the infinite space-time volume.

%% file: sec3.tex
\section{Finite $V$ and fixed $Q$ effects 
within ChPT}
\label{sec:ChPT}

In this section, we first discuss how to evaluate 
the effect of fixing topology.
A general argument leads to a consequence that 
the dependence on the global topological charge 
only appears with a suppression factor $1/V$. 
Namely, it is a part of the finite volume effects.
Recent studies of the finite volume scaling
within ChPT are then reviewed.
Once we assume that the heavier hadrons, such as
rho mesons, baryons {\it etc}, are all decoupled from
the theory at the scale of $1/V^{1/4}$, 
only pions describe the difference of the 
finite volume system from the infinite volume one.  
We discuss, in particular,
a non-perturbative approach to integrate over the chiral 
field's vacuum, which is necessary in the $\epsilon$-regime.

\subsection{Topology as an infra-red physics}
Let us start our discussion with an intuitively noticeable difference 
between the trivial topological sector ($Q=0$) and
the first non-trivial one ($Q=1$).
In the weak coupling limit $g\ll 1$, it is well-known that a self-dual
solution,  the so-called one-instanton solution, dominates
the configuration space of the $Q=1$ sector 
and its relative weight is given by $\sim \exp(-8\pi^2/g^2)$.
For larger value of $Q$, the weight is expected to be
$\sim \exp(-8\pi^2|Q|/g^2)$.
As the coupling constant becomes strong,
$g\sim 1$, more complicated configurations with 
many pairs of instantons and anti-instantons 
are more favored, since the entropy gives more impact
on the free energy than the action density.
Suppose that the number of such pairs generated 
in a typical configuration is $Q_{ave}$.
The trivial sector $Q=0$ then has
$Q_{ave}$ instantons and $Q_{ave}$ anti-instantons
while in the $Q=1$ sector $(Q_{ave}+1)$ instantons 
and $Q_{ave}$ instantons are there.
As $Q_{ave}$ grows, the difference between the global topological charge,
$Q=0$ and $Q=1$ would become less important.

If the theory has a mass gap $\Lambda_{gap}$
(it is natural to assume $\Lambda_{gap}=\Lambda_{QCD}$ for 
the pure gauge theory while $\Lambda_{gap}$ is the pion mass $m_\pi$ for QCD), 
the typical size of an instanton or anti-instanton
should be given by $1/\Lambda_{gap}$ and their 
density is estimated as $\sim \Lambda_{gap}^4$.
The value of $Q_{ave}$ discussed above 
is then estimated by $\sim \Lambda_{gap}^4 V$
and one can easily see how the 
difference between $Q=0$ and $Q=1$ (or higher)
disappears as $\sim 1/V$ when $V$ is sent to infinity
or equivalently $Q_{ave}\to \infty$.
The effect of the global topological charge thus
should be understood as a finite volume effect. 

Brower et al. \cite{Brower:2003yx}  and Aoki et al. \cite{Aoki:2007ka}
gave a more theoretical and solid formulation for the effect of the
global topological charge. 
The partition function of the theory at a fixed topological
charge $Q$ is obtained from those at the $\theta$ vacua 
by a Fourier transformation
\begin{eqnarray}
Z_Q &=& \int d\theta\; e^{i\theta Q}Z(\theta) = \int d\theta\; 
e^{i\theta Q}\exp(-f(\theta)V),
\end{eqnarray}
where $f(\theta)$ denotes a free-energy density of the $\theta$ vacuum.
When the vacuum angle $\theta$ is small,
$f(\theta)$ can be expanded in $\theta^2$ as \cite{Vafa:1984xg}
\begin{eqnarray}
f(\theta) &=& \frac{\chi_t}{2}\theta^2 + c_4 \theta^4 +c_6 \theta^6+ \cdots,
\end{eqnarray}
where a constant term is omitted.
Here, $\chi_t$ corresponds to the topological susceptibility.
Assuming that all the constants, $\chi_t$, $c_4$, $c_6$ {\it etc}.
are of the order of $\sim (\Lambda_{gap})^4$
and the volume is large enough to satisfy $L \Lambda_{gap}\gg 1$, 
the above $\theta$ integral can be
evaluated by a saddle-point expansion as
\begin{eqnarray}
Z_Q &=& \frac{1}{\sqrt{2\pi \chi_t V}}
\exp\left(-\frac{Q^2}{2\chi_t V}\right)
\left[1-\frac{c_4}{8\chi_t^2 V}+\cdots\right],
\end{eqnarray}
which clearly shows that the global topological charge dependence 
disappears in the limit $V\to\infty$.
It is also important to notice that the distribution of 
the global topological charge converges
to the Gaussian distribution
as the volume increases, which agrees well with 
the intuitive picture above that only the entropy given by
the distribution of instantons and anti-instantons becomes
important in the thermodynamical limit.

Under those minimal assumptions on the vacuum free energy, one can
prove that $\chi_t$, $c_4$ {\it etc}. can be extracted from lattice
QCD simulations at a fixed topological charge \cite{Aoki:2007ka}.
For instance, $\chi_t$ appears as a constant mode in the two-point
correlator in the flavor singlet channel 
\begin{eqnarray}
\langle \eta^\prime (x)\eta^\prime (y)\rangle_Q
&=& -\frac{\chi_t}{V} + {\cal O}(1/V^2)+{\cal O}(e^{-m_{\eta^\prime}|x-y|}),
\end{eqnarray}
for a large separation $|x-y|$. 
The excitation in this channel corresponds to the $\eta'$
meson whose non-zero mass is given by
$m_{\eta^\prime}$.
The constant correlation has a negative sign when the global
topological charge $Q$ is zero, because at long distances there is
more chance to find oppositely charged local topological excitations
when the sum is constrained to zero.

\begin{figure}[tbp]
  \centering
  \includegraphics[width=10cm,clip=true]{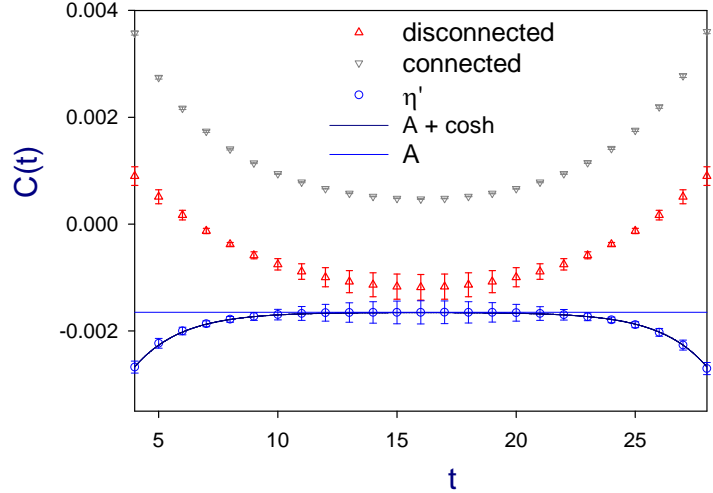}
  \caption{
The eta-prime correlator (circles) 
at $m=0.002$ and $Q=0$ obtained in the two-flavor QCD simulation.
A negative constant contribution is seen.
The triangles are its connected and disconnected diagram parts.
    }
  \label{fig:etaprime}
\end{figure}

\begin{figure}[tbp]
  \centering
  \includegraphics[width=10cm,clip=true]{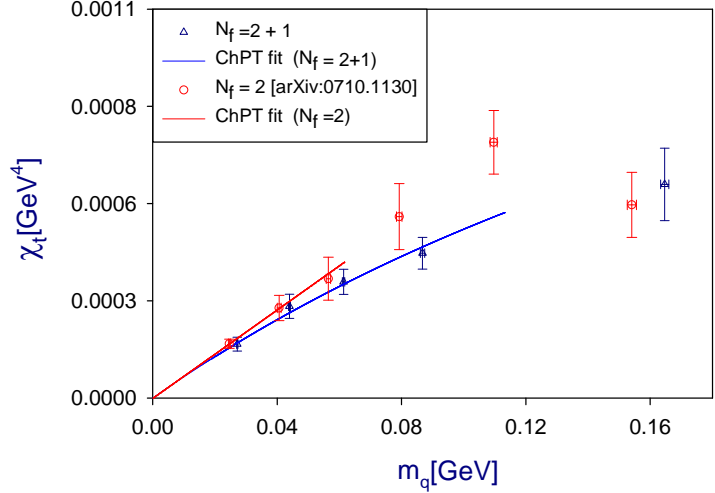}
  \caption{
    $\chi_t$ extracted from the $\eta'$ meson correlators.
    A good agreement with ChPT predictions (solid lines)
    is seen both in the  $N_f=2$ and $N_f=2+1$ lattice data.
    }
  \label{fig:chit}
\end{figure}

In the numerical simulations \cite{Aoki:2007pw, Chiu:2008kt, Chiu:2008jq, Hsieh}
the presence of this constant mode is confirmed
as Fig.~\ref{fig:etaprime} shows.
Moreover, the extracted values of $\chi_t$ via above formula
are found to agree with the ChPT prediction
\cite{Leutwyler:1992yt} 
\begin{equation}
\chi_t = \frac{\Sigma}{\sum_f^{N_f}1/m_f},
\end{equation}
as seen in Fig.~\ref{fig:chit}. 
The value of $\Sigma$ extracted from this analysis is consistent with
a nominal value $\Sigma\simeq$ (250~MeV)$^3$.
Chiral fit including the next-to-leading chiral corrections 
\cite{Mao:2009sy,Aoki:2009mx} is underway.

There are two remarkable conclusions that may be drawn from
these lattice data. 
First, local fluctuation of topology 
exists even when the global topological charge is fixed 
in Monte Carlo simulations.
There was some doubt about the ergodicity of the Monte Carlo
simulation with the topology fixing term, but as far as the numerical
data imply there is no evidence of the problem.
Second, the topological charge actually feels the presence of
dynamical fermions and the $\chi_t$ vanishes in the chiral limit as
expected from ChPT. 
Topology is a part of the infrared physics that can
be well described by the pion physics.

\subsection{Finite $V$ and fixed $Q$ within ChPT}

The Lagrangian of ChPT is given by \cite{Gasser:1983yg}
\begin{eqnarray}
\mathcal{L}&=& \frac{F^2}{4}
{\rm Tr}[\partial_\mu U(x)^\dagger \partial_\mu U(x)]
-\frac{\Sigma}{2}{\rm Tr}
[\mathcal{M}^\dagger e^{-i\theta/N_f}U(x)
+U(x)^\dagger e^{i\theta/N_f}\mathcal{M}]+\cdots,
\end{eqnarray}
where the chiral field $U(x)$ is an element of $SU(N_f)$ group.
Here the pion decay constant and the chiral condensate are
denoted by $F$ and $\Sigma$, respectively.
The vacuum angle $\theta$ is given as a phase in front of the
mass matrix $\mathcal{M}=\mbox{diag}(m_u, m_d, m_s, \cdots)$.

In the conventional $p$-expansion, we treat 
the exponent of $U(x)$ as the Nambu-Goldstone modes (here we denote as
$\xi(x)$), or pions, and expand the chiral field as
\begin{eqnarray}
U(x) &=& \exp\left(i\frac{\sqrt{2}\xi(x)}{F}\right)=
1+i\frac{\sqrt{2}}{F}\xi(x)-\frac{1}{F^2}\xi^2(x)+\cdots.
\end{eqnarray}
With the counting rule 
\begin{eqnarray}
\label{eq:pcounting}
\mathcal{M}\sim p^2,\;\;\; \partial_\mu \sim p,\;\;\;
1/L,  1/T \sim p,\;\;\; \xi(x) \sim p,
\end{eqnarray}
physical amplitudes are systematically expanded in terms of $p^2$.

In the $p$-regime,
the finite volume effect appears in the pion propagator since
the momentum space is discretized \cite{Bernard:2001yj}.
Pion correlator reads
\begin{eqnarray}
\label{eq:pcorr}
\langle \xi^a(x)\xi^b(y)\rangle = \delta_{ab}
\sum_p \frac{e^{ip(x-y)}}{p^2+m_\pi^2},
\end{eqnarray}
where $a$($b$) denotes the $a$($b$)-th generator of $SU(N_f)$
and the summation is taken over the 4-momentum 
$
p=2\pi(n_t/T, n_x/L, n_y/L, n_z/L),
$
with integer $n_\mu$'s.
As a consequence, 
all the correlators become periodic.
Even at a contact point
$x=y$, there exists a finite volume correction
\begin{eqnarray}
\label{eq:g1}
\langle \xi^a(x)\xi^b(x)\rangle &=&\delta_{ab}\left(
\frac{m_\pi^2}{16\pi^2}\ln \frac{m_\pi^2}{\mu_{sub}^2}
+g_1(m_\pi^2)\right),\\
g_1(M^2)&=& \sum_{a\neq 0}\frac{M}{4\pi^2|a|}K_1(M|a|), 
\end{eqnarray}
which is understood as an effect of pion wrapping around the lattice. 
Here $K_1(x)$ is the modified Bessel function and the summation
is taken over the 4-vector $a_\mu=n_\mu L_\mu$ with 
$L_i=L$ for $i=1,2,3$ and $L_4=T$,
except for $a_\mu=(0,0,0,0)$.
Note that the subtraction of the ultraviolet divergence is done
at a scale $\mu_{sub}$, which can be made in exactly the same way 
as in the infinite volume.
In a similar perturbative manner, the effect of global topology 
is recently calculated to the next-to-leading order \cite{Aoki:2009mx}.

In the $\epsilon$-regime, the above $p$-expansion
(\ref{eq:pcounting})
fails because the zero-momentum mode contribution
induces an unphysical infrared divergence,
which has to be circumvented
by exactly treating the vacuum fluctuation of the chiral field. 
Namely, using a parameterization
\begin{eqnarray}
U(x) &=& U_0 \exp\left(i\frac{\sqrt{2}\xi^\prime(x)}{F}\right),
\end{eqnarray}
where $U_0\in SU(N_f)$ and $\xi^\prime$ satisfies
\begin{eqnarray}
\int d^4 x\; \xi^\prime(x) &=&0,
\end{eqnarray}
one can explicitly factorize the zero momentum part as $U_0$.
Since  $U_0$ has no dependence on $x$,
the group integral can be non-perturbatively performed
as in the calculation of random matrix models. 
The non-zero momentum modes $\xi^\prime$'s 
are perturbatively treated as an expansion in $\epsilon^2$ according
to the counting rule 
\begin{eqnarray}
\mathcal{M}\sim \epsilon^4,\;\;\; \partial_\mu \sim \epsilon,\;\;\;
1/L,  1/T \sim \epsilon,\;\;\; \xi'(x) \sim \epsilon. 
\end{eqnarray}
This $\epsilon$-expansion \cite{Gasser:1987ah, Leutwyler:1992yt}
is useful when the quark mass is 
so small that the pion correlation length exceeds
the spatial extent, $m_\pi L \ll 1$.

The zero momentum component $U_0$ can be explicitly integrated out and
written in terms of analytic functions.
This fact opens an interesting theoretical opportunities.
In particular, at the leading order of the $\epsilon$-expansion, the
system is proven to be equivalent to the Random Matrix Theory
\cite{Shuryak:1992pi,Damgaard:1998xy,Basile:2007ki}.
In the context of the QCD study, this provides a new method to
determine the chiral condensate by matching the low-lying eigenvalues
of the Dirac operator.
At an early stage, a simple setup with all degenerate quark masses
were studied. 
As lattice QCD is developed to reach the simulations near the
chiral limit, calculations in a more realistic setup has become
relevant, and partially quenched calculations of various quantities
have been carried out \cite{Bernardoni:2007hi,Damgaard:2000di}.
With strange quark mass kept at its physical value, the finite volume
system is not purely in the $\epsilon$-regime even when the up and
down quark masses are sent close to the chiral limit, because the kaon
and $\eta$ are heavy and do not satisfy $m_{K,\eta} L \ll 1$.
For this mixed-regime, a hybrid method is proposed
\cite{Bernardoni:2008ei} and even extended to the case of heavy-light
mesons \cite{Bernardoni}. 
More recently, a theoretical framework in which the $\epsilon$- and
$p$-regimes are treated in a unified manner is proposed
\cite{Damgaard:2008zs} of which 
details are described in the next section.

As a final remark of this section, we note the role of topological
charge in the $\epsilon$-regime.
Intuitively, the global topological charge become relevant to the
dynamics of the system when the volume is small.
This can be explicitly studied within ChPT.
The $\theta$ integral can be absorbed in the zero-mode integrals
\begin{eqnarray}
\int \frac{d\theta}{2\pi}\int_{SU(N_f)} dU_0 e^{i\theta Q}
=\int_{U(N_f)} d U_0 (\det U_0)^Q,
\end{eqnarray}
where $U_0$ is integrated over U($N_f$) manifold.
The effect of the topological charge
enters through a factor $(\det U_0)^Q$.
For instance, the spectral density of low-lying Dirac eigenmodes is
largely affected by $Q$, of which dependence can be used
to test the validity of ChPT, in addition to the quark mass dependence.

%% file: sec4.tex
\section{Determination of the chiral condensate}
\label{sec:cond}

\subsection{Analytic results beyond the leading order}
The chiral condensate is related to the Dirac eigenvalue density
$\rho(\lambda)$ at $\lambda=0$ in the thermodynamical limit
\cite{Banks:1979yr} as $\rho(0)=\Sigma/\pi$.
This relation can be easily extended to non-zero eigenvalues
by an analytical continuation of the valence mass $m_v$
to a pure imaginary value $i\lambda$:
\begin{eqnarray}
\rho(\lambda) &=& \frac{1}{\pi}{\rm Re}\langle \bar{q_v}q_v \rangle |_{m_v=i\lambda}.
\end{eqnarray}
Here, $\bar{q}_vq_v$ is the scalar density operator made of the
valence quark field.
This general formula is valid for both $p$- and $\epsilon$-regimes.

In the $p$-regime, using the partial quenching technique for the
imaginary valence quark mass, 
Osborn {\it et al.} \cite{Osborn:1998qb} (see also \cite{Smilga:1993in}) 
found that the Dirac spectrum contains a logarithmic dependence on
$\lambda$.
This calculation is done in the infinite volume limit with degenerate
quark masses.

For small eigenvalues, the effect of finite volume becomes important.
The ChPT calculation is simplified if one consider the
$\epsilon$-expansion and taking its leading order contribution.
The integral over the zero momentum pion mode can be done
analytically, and the spectral function has been obtained as a
function of $N_f$, sea quark masses, and topological charge $Q$
\cite{Damgaard:2000ah,Verbaarschot:1993pm,Akemann:1998ta}.
Except for the exact zero-modes associated with $Q$, there is 
a finite gap from zero (of order $1/\Sigma V$, which is called the
microscopic region) in the Dirac operator spectrum. 
These analytic ChPT results can be used to extract $\Sigma$ by
comparing with the lattice data in the $\epsilon$-regime
\cite{Fukaya:2007fb,DeGrand:2006nv}.
But, since the formulae are obtained at the leading order, the value
of $\Sigma$ thus obtained is a subject of the NLO corrections of the
$\epsilon$-expansion. 
Furthermore, it requires that the system is in the $\epsilon$-regime,
which is numerically demanding.
For common lattice QCD configurations produced in a $p$-regime
set-up, these analytical results cannot be applied.

Here we introduce a new method of the chiral expansion
\cite{Damgaard:2008zs}.
It is based on the $p$-expansion, but includes the pion zero-mode
integral explicitly so that a transition to the $\epsilon$-regime is
smooth. 
In this scheme, one may predict the eigenvalue spectrum in the
microscopic region for the system in the $p$-regime.
With the so-called replica trick, the calculation is extended to the 
case of non-degenerate quarks of arbitrary number of flavors.

At a fixed topological charge $Q$, we obtain \cite{Damgaard:2008zs}
\begin{eqnarray}
\label{eq:rho}
\rho_Q(\lambda) &=& \Sigma_{\rm eff}
\hat{\rho}^\epsilon_Q(\lambda \Sigma_{\rm eff}V,\{m_{sea}\Sigma_{\rm eff}V\})
+\rho^p(\lambda,\{m_{sea}\}),
\end{eqnarray}
where $\lambda$ denotes the Dirac eigenvalue,
$\{m_{sea}\Sigma_{\rm eff}V\}=\{m_1\Sigma_{\rm eff}V,m_2\Sigma_{\rm eff}V,\cdots\}$
is a set of the sea quark masses normalized by
an {\it effective} chiral condensate $\Sigma_{\rm eff}$ 
(the definition is given below) and $V$.
The first term on the right hand side of (\ref{eq:rho}) contains
the one in the leading-order $\epsilon$-expansion 
$\hat{\rho}^\epsilon_Q(\zeta,\{\mu_{sea}\}=\{\mu_1,\mu_2\cdots\})$,
which is rescaled so that the physical scale $\Sigma$ is factored
out. 
This is a known function given by determinants of the Bessel functions
\cite{Akemann:1998ta}: 
\begin{eqnarray}
\hat{\rho}^\epsilon_Q(\zeta,\{\mu_{sea}\})
\equiv 
C_2 \frac{|\zeta|}{2\prod^{N_f}_f(\zeta^2 + \mu^2_f)}
\frac{\det \tilde{\mathcal{B}}}{\det \mathcal{A}},
\end{eqnarray}
where an $N_f\times N_f$ matrix $\mathcal{A}$ and an
$(N_f+2)\times (N_f+2)$ matrix $\tilde{\mathcal{B}}$ are defined by
\begin{eqnarray}
\mathcal{A}_{ij}&=& \mu_i^{j-1}I_{Q+j-1}(\mu_i),\\
\tilde{\mathcal{B}}_{1j} &=&  \zeta^{j-2}J_{Q+j-2}(\zeta),\;\;\;
\tilde{\mathcal{B}}_{2j} =  \zeta^{j-1}J_{Q+j-1}(\zeta),\nonumber\\
\tilde{\mathcal{B}}_{ij} &=&  (-\mu_{i-2})^{j-1}I_{Q+j-1}(\mu_{i-2})
\;\;\;(i\neq 1,2).
\end{eqnarray}
Here, the overall sign is $C_2=+1$ for the $N_f=2$ and 3 cases.

The second term in (\ref{eq:rho}) is a logarithmic
NLO correction as always seen in the conventional $p$-expansion.
Defining $M_{ij}^2\equiv (m_i+m_j)\Sigma/F^2$, the function is given by
\begin{eqnarray}
\label{eq:rhop}
\rho^p(\lambda,\{m_{sea}\}) &\equiv&
-\frac{\Sigma}{\pi F^2}{\rm Re}[
\sum^{N_f}_f (\bar{\Delta}(M^2_{fv})
-\bar{\Delta}(M^2_{ff}/2))
\left.-(\bar{G}(M^2_{vv})-\bar{G}(0))
\right]_{m_v=i\lambda},
\end{eqnarray}
where 
\begin{eqnarray}
\bar{G}(M^2) &=& \left\{
\begin{array}{l}
\frac{1}{2}\left[ 
\bar{\Delta}(M^2) +
(M^2-M_{ud}^2)\partial_{M^2}\bar{\Delta}(0,M^2)\right] 
\hspace{1.4in}(N_f=2),\\
\frac{1}{3}\left[ 
-\frac{2(M^2_{ud}-M^2_{ss})^2}
{9(M^2-M^2_{\eta})^2}\bar{\Delta}(M_\eta^2)
\;\;\;+\left(1 + \frac{2(M_{ud}^2-M_{ss}^2)^2}{9(M^2-M^2_{\eta})^2}\right)
\bar{\Delta}(M^2)
\right.\\\left.
\;\;\;
+\frac{(M^2-M^2_{ud})(M^2-M^2_{ss})}{(M^2-M^2_{\eta})}
\partial_{M^2}\bar{\Delta}(M^2)\right] 
\hspace{1.5in}
(N_f=2+1),\\
\end{array}\right.\nonumber\\\\
\bar{\Delta}(M^2)&=& 
\frac{M^2}{16\pi^2}\ln \frac{M^2}{\mu_{sub}^2}
+\bar{g}_1(M^2).
\label{eq:Deltabar}
\end{eqnarray}
Here 
$M_{ud}^2=2m_u\Sigma/F^2=2m_d\Sigma/F^2$, $M_{ss}^2=2m_s\Sigma/F^2$
and $M_\eta^2=(M_{ud}^2+2M_{ss}^2)/3$.
The function $\bar{g}_1(M^2)=g_1(M^2)-1/M^2V$ denotes
the well-known finite volume correction from non-zero modes
\cite{Bernard:2001yj}
(see also (\ref{eq:g1})). 
The scale $\mu_{sub}^2$(=770 MeV in this work) is a subtraction scale.
Note that $\rho^p(\lambda,\{m_{sea}\}$ is insensitive to the
topological charge.

The {\it effective}
condensate in (\ref{eq:rho}) is also expressed in terms of 
$\bar{\Delta}(M^2)$ and $\bar{G}(M^2)$ as
\begin{eqnarray}
\label{eq:Sigmaeff}
\Sigma_{\rm eff} &\equiv & \Sigma \left[1-\frac{1}{F^2}\left(
\sum^{N_f}_f \bar{\Delta}(M^2_{ff}/2)-\bar{G}(0)
-16L^r_6\sum^{N_f}_f M^2_{ff}\right)\right].
\end{eqnarray}
This depends on the sea quark masses, volume $V$ and $L_6^r$ of which value is
renormalized (at $\mu_{sub}=770$ MeV in this work).

\begin{figure}[tbp]
  \centering
  \includegraphics[width=10cm,clip=true]{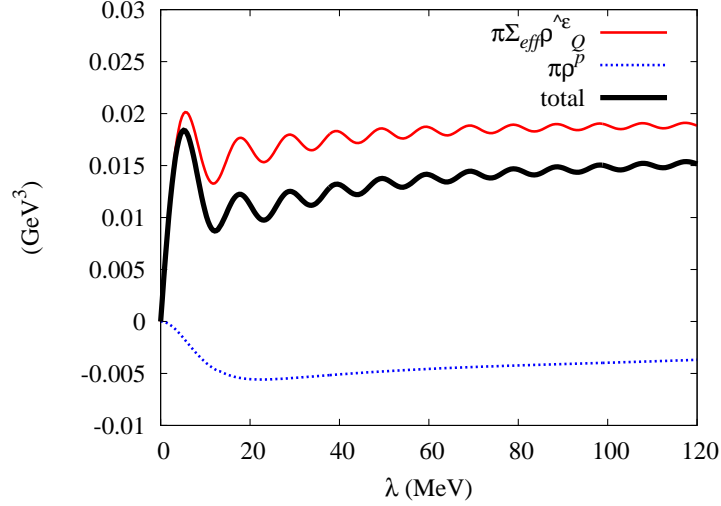}
  \caption{
    The first term $\Sigma_{\rm eff}\hat{\rho}_Q^\epsilon$ (solid-thin curve),  
    the second term $\rho^p$ (dashed)
    and the total contribution (solid-thick) 
    of the spectral density (\protect\ref{eq:rho}) are shown.
    The curves are multiplied by $\pi$.
    We use $\Sigma=[240\mbox{MeV}]^3$, $F=94$ MeV, $L_6^r=-0.0001$, $L=T/3=1.9$ fm,
    $m_{ud}=20$ MeV and $m_s=120$ MeV as inputs.
    }
  \label{fig:rhoChPT}
\end{figure}

For an illustration,
we draw curves given by the formula (\ref{eq:rho})
in Fig.~\ref{fig:rhoChPT}.
The contributions from 
the first term $\Sigma_{\rm eff}\hat{\rho}^\epsilon_Q$ (solid-thin curve),
the second term $\rho^p$ (dashed) 
and the total contribution $\rho_Q(\lambda)$ (solid-thick)
are shown separately.
We use typical parameters
$\Sigma=[240\mbox{MeV}]^3$, 
$F$ = 94~MeV, 
$L_6^r=-0.0001$, 
$L=T/2$ = 1.9~fm,
$m_{ud}$ = 20~MeV and 
$m_s$ = 120~MeV as inputs.
One can see that the second term gives a negative contribution and
shows a significant curvature in the lower end of the spectrum. 
This is the effect of the chiral logarithm.
For this quark mass, the formula starts to deviate from 
the leading order expression in the $\epsilon$-expansion 
already at $\lambda\sim 5$ MeV.

\subsection{A numerical analysis}
Our simulation details and parameters have been
already presented in Section.~\ref{sec:ov}.
For the study of the Dirac spectrum, 
80 lowest pairs of eigenvalues of the overlap-Dirac
operator $D(0)$ are calculated at every 5--10 trajectories.
We employ the implicitly restarted Lanczos algorithm for the
chirally projected operator $P_+\,D(0)\,P_+$, where
$P_+\!=\!(1+\gamma_5)/2$. 
From its eigenvalue ${\rm Re} \lambda^{ov}$,
the pair of eigenvalues $\lambda^{ov}$ (and its complex conjugate) 
of $D(0)$ is extracted through the relation 
$|1-\lambda^{ov}/m_0|^2=1$, that forms a circle on a complex plane. 
For the comparison with the effective theory, 
the lattice eigenvalue
$\lambda^{ov}$ is projected onto the imaginary axis as
$\lambda\equiv\mathrm{Im}\lambda^{ov}/(1-\mathrm{Re}\lambda^{ov}/(2m_0))$.
Note that the real part of $\lambda^{ov}$ is negligible 
(within 1\%) for the low-lying modes.

When we match the lattice data for the spectral density with the
analytic calculation (\ref{eq:rho}), 
two parameters are to be determined at each set of 
the quark masses: $\Sigma_{\rm eff}$ and $F$.
In the second NLO term of (\ref{eq:rho}), the difference between 
$\Sigma_{\rm eff}$ and $\Sigma$ is a higher order effect.
We therefore take two reference values of $\lambda$ to give inputs to
determine $\Sigma_{\rm eff}$ and $F$.
The reference points are chosen such that they have maximum
sensitivity to the parameters  
in the convergence range of the chiral expansion:
$\lambda =0.004$ ($\sim 7$ MeV) and 0.017 ($\sim 30$ MeV) 
except for the case with $m_{ud}=0.002$ and $Q=1$,
for which we choose $\lambda=0.01$ and 0.02 
(because of its weaker sensitivity to the NLO effects).
At these two reference points, we compare the mode number
below a given value of $\lambda$ \cite{Giusti:2008vb},
with an integrated formula of ChPT (\ref{eq:rho})
\begin{eqnarray}
\label{eq:cum}
N_Q(\lambda) &\equiv& V\int^{\lambda}_0 d\lambda^\prime \rho_Q (\lambda^\prime),
\end{eqnarray}
and determine $\Sigma_{\rm eff}$ and $F$.
As Giusti and L\"uscher \cite{Giusti:2008vb} studied, 
it is also useful to define a quantity
\begin{eqnarray}
\label{eq:Sigmamode}
\Sigma^{mode}_Q(\lambda) 
&\equiv& \frac{\pi N_Q(\lambda)}{\lambda V},
\end{eqnarray}
to see the NLO effects, or the chiral 
logarithmic effects to $\Sigma$.
We test the both of $N_f=2+1$ and $N_f=2$ ChPT formulae.
For the latter case, the strange quark is assumed 
to be decoupled from the theory.

\begin{figure*}[tbp]
  \centering
  \includegraphics[width=9.5cm]{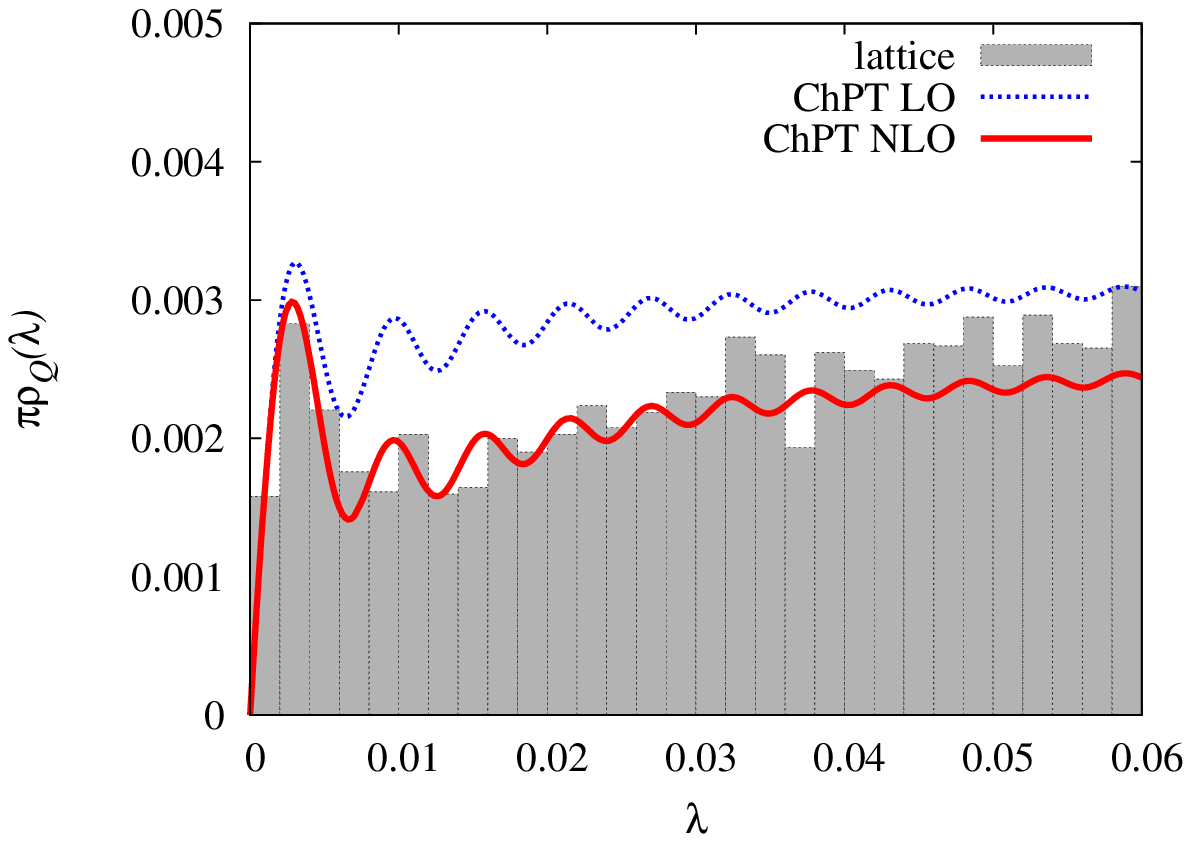}
  \includegraphics[width=9.5cm]{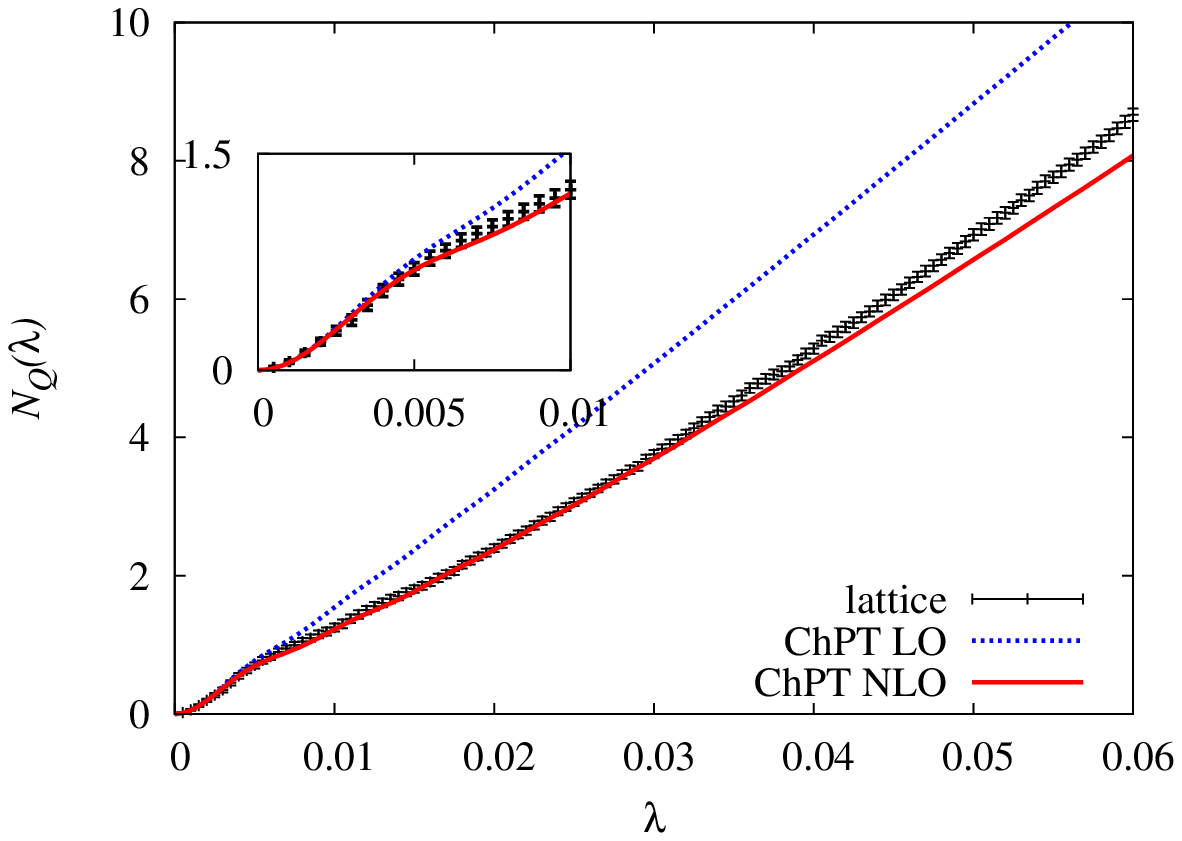}
  \includegraphics[width=9.5cm]{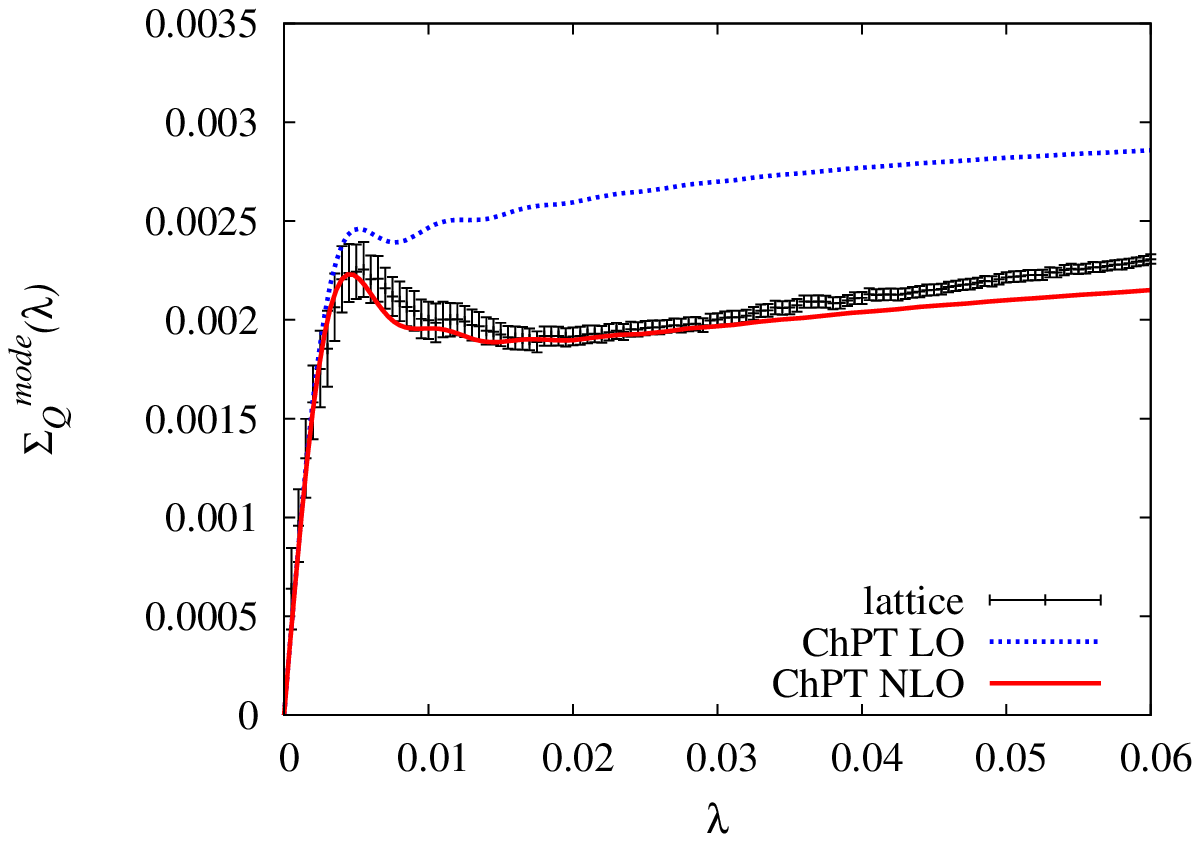}
  \caption{
    The spectral density $\pi\rho_Q(\lambda)$ (top), the mode 
    number $N_Q(\lambda)$ (center) and $\Sigma^{mode}_Q(\lambda)$ (bottom)
    of the Dirac operator at $m_{ud}$ = 0.015, $m_s$ = 0.080 and $Q=0$.
    The lattice result (histogram (top) or solid symbols (center and bottom)) is
    compared with the ChPT formula drawn by solid curves.
    For comparison, the prediction of the leading
    $\epsilon$-expansion (dashed curves) is also shown.
  }
  \label{fig:rho015}
\end{figure*}

\begin{figure*}[tbp]
  \centering
  \includegraphics[width=9.5cm]{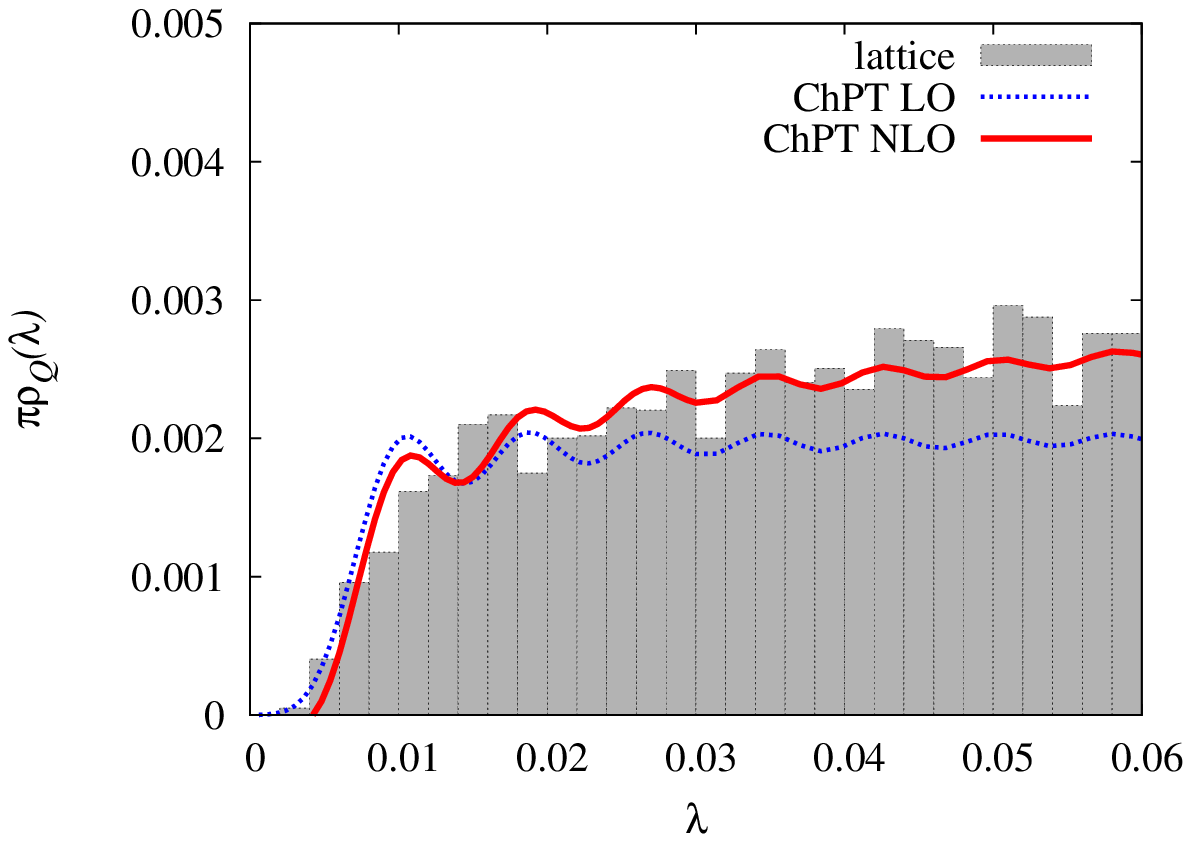}
  \includegraphics[width=9.5cm]{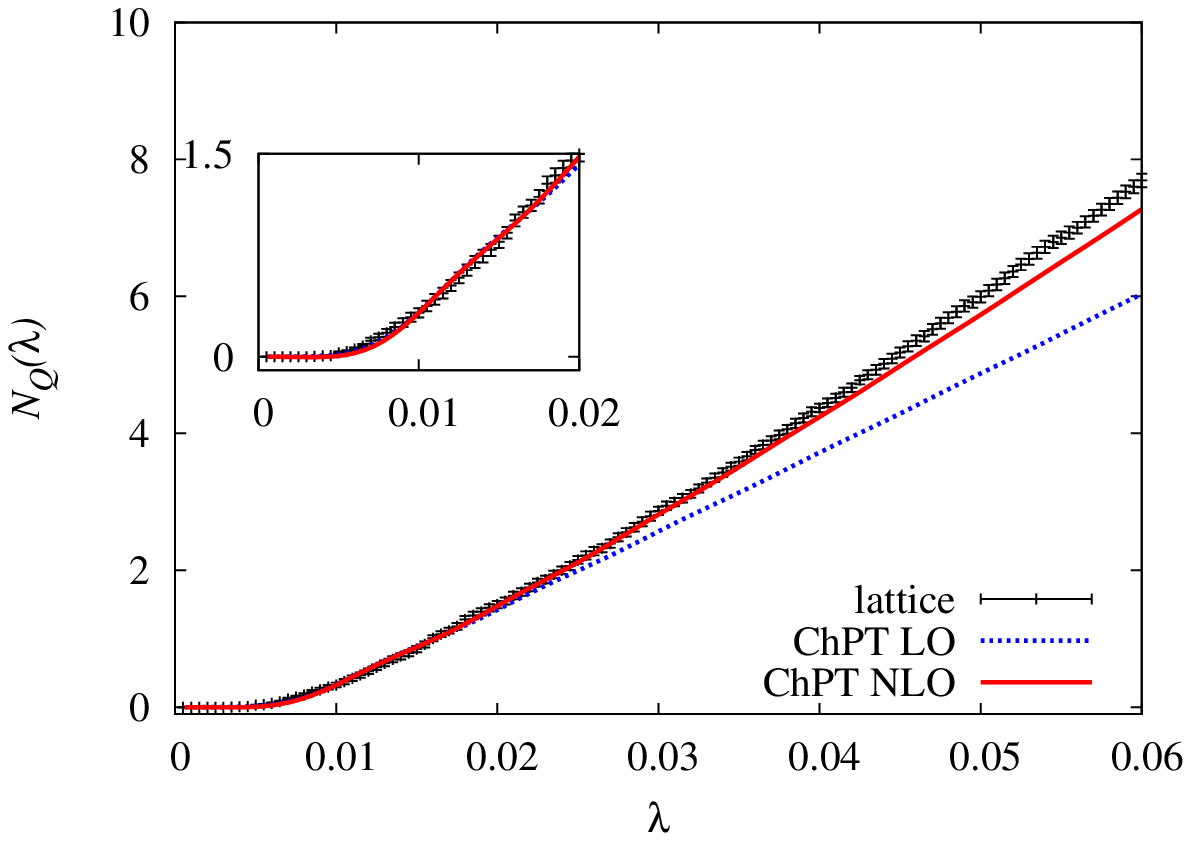}
  \includegraphics[width=9.5cm]{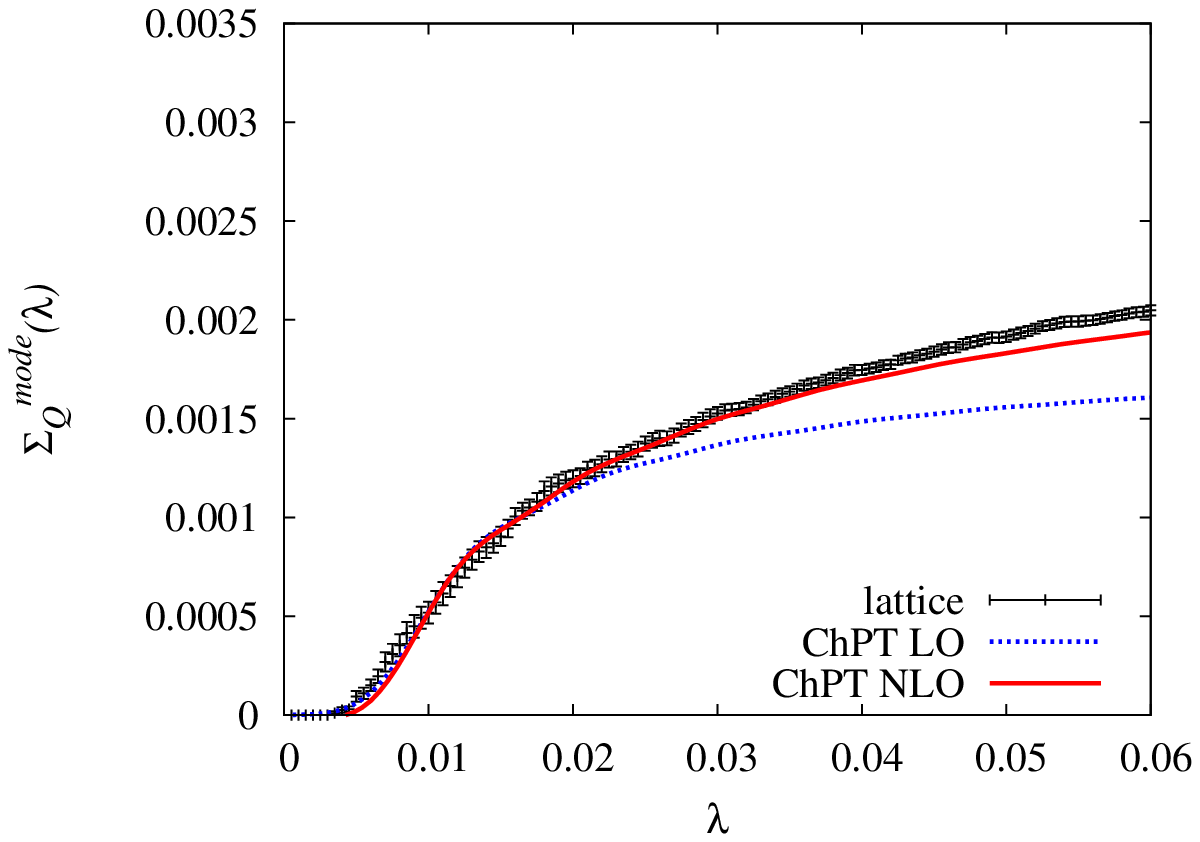}
  \caption{
    Same as Fig.~\protect\ref{fig:rho015}, but at
    $m_{ud}=0.002$. 
    The NLO correction is smaller in the $\epsilon$-regime.
  }
  \label{fig:rho002}
\end{figure*}

Figures~\ref{fig:rho015} and \ref{fig:rho002} show the lattice data
for the spectral density (upper panel), its integral (middle)
and $\Sigma^{mode}_Q(\lambda)$ defined by (\ref{eq:Sigmamode}) 
at two different sea quark masses:
one in the $p$-regime ($m$ = 0.015, Fig.~\ref{fig:rho015}) and the
other in the $\epsilon$-regime ($m$ = 0.002, Fig.~\ref{fig:rho002}).
The analytic formula is also plotted with two parameters fixed at two
reference points of the mode number.
The leading-order contribution is given by dotted curves while the
full result is shown by solid curves.

In the $p$-regime result (Fig.~\ref{fig:rho015}),
the effect of the NLO term in the $p$-expansion is clearly seen as a
deviation from the leading-order density
$\Sigma_{\rm eff}\hat{\rho}_Q^\epsilon$
(dotted curve) in the histogram.
The deviation starting already around $\lambda \sim 0.005$
is also clear in the mode number $N_Q(\lambda)$ and $\Sigma_Q^{mode}(\lambda)$.
On the other hand, the NLO formula (solid curve) describes the lattice
data very nicely up to $\lambda\sim m_s/2$.

The convergence of the chiral expansion is better for the
$\epsilon$-regime data (Fig.~\ref{fig:rho002}), but the difference
between LO and NLO still exists.
We also observe that there is a wider gap near $\lambda=0$, which is
expected because the value of the sea quark mass $m=0.002$ is similar
to the lowest eigenvalue, so that the suppression due to the fermionic
determinant $\prod(\lambda^2+m^2)$ works strongly.

\begin{figure*}[tbp]
  \centering
  \includegraphics[width=10cm]{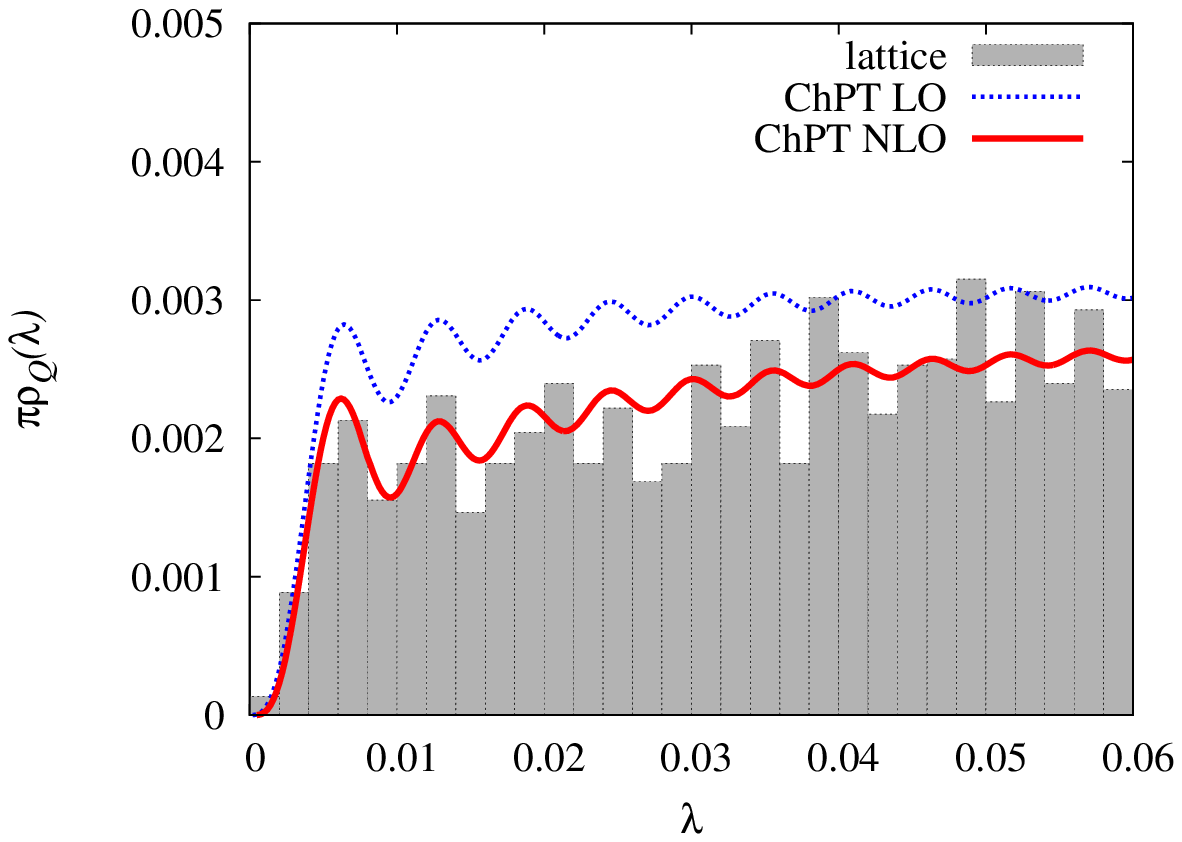}
  \includegraphics[width=10cm]{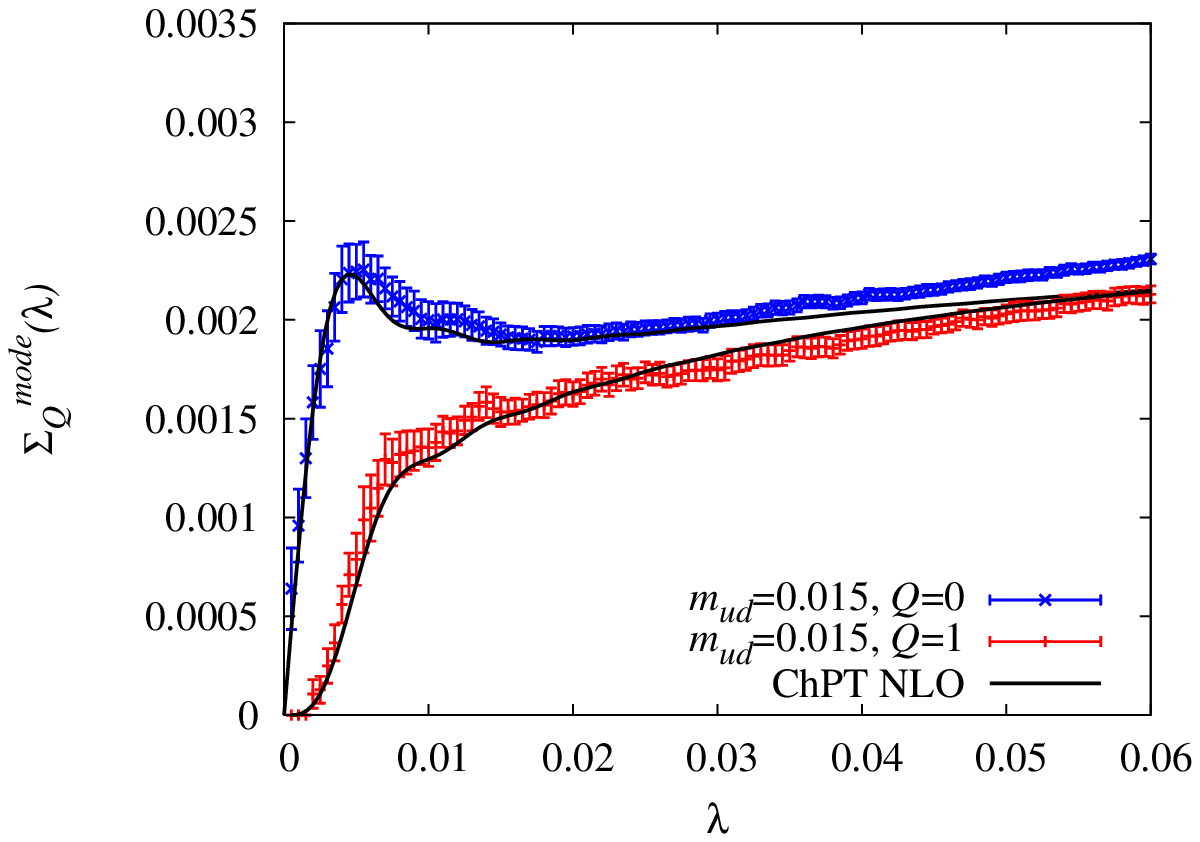}
  \caption{
    The spectral density at $m=0.015$ and $Q=1$ (top)
    and comparison of $\Sigma_Q^{mode}(\lambda)$ at $Q=0$ and 1 (bottom). 
    In the ChPT curves, the same values of $\Sigma_{eff}$ and $F$ 
    are used as inputs.
  }
  \label{fig:rhoQ}
\end{figure*}

\begin{figure*}[tbp]
  \centering
  \includegraphics[width=10cm]{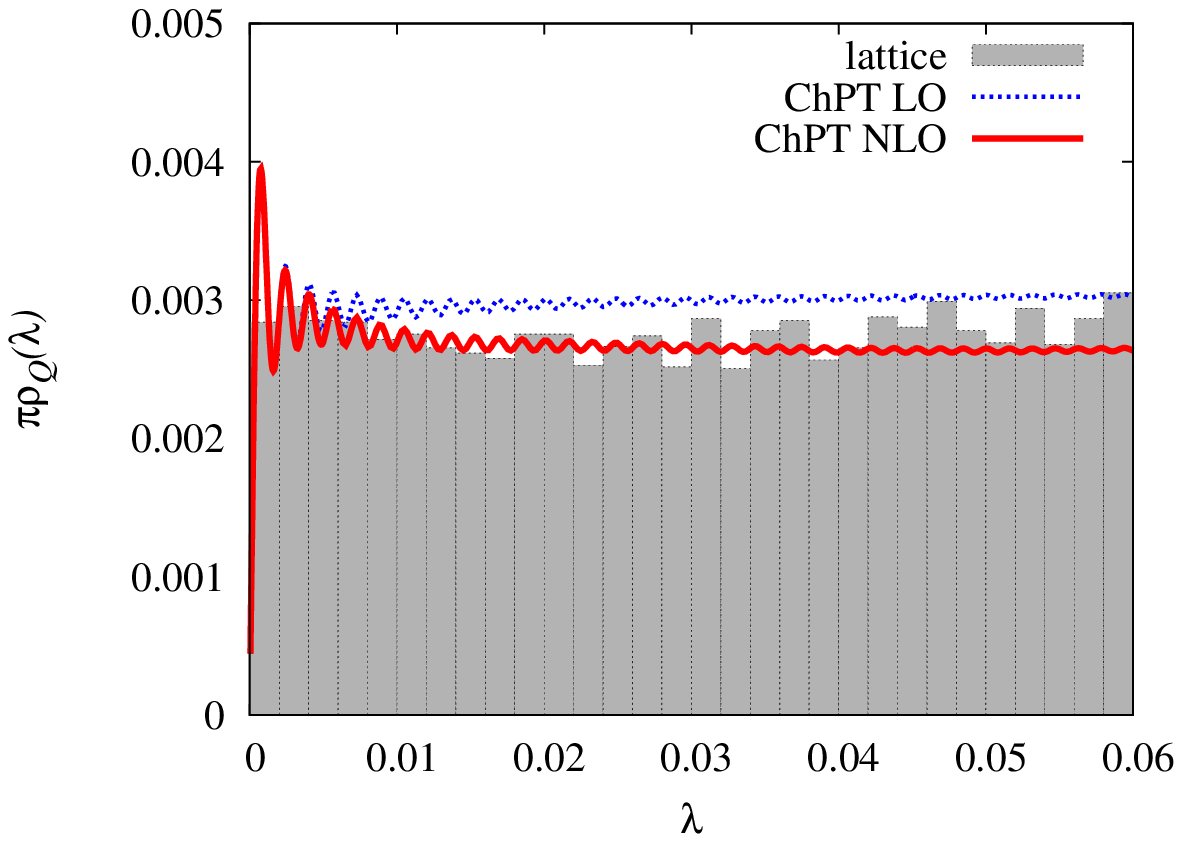}
  \includegraphics[width=10cm]{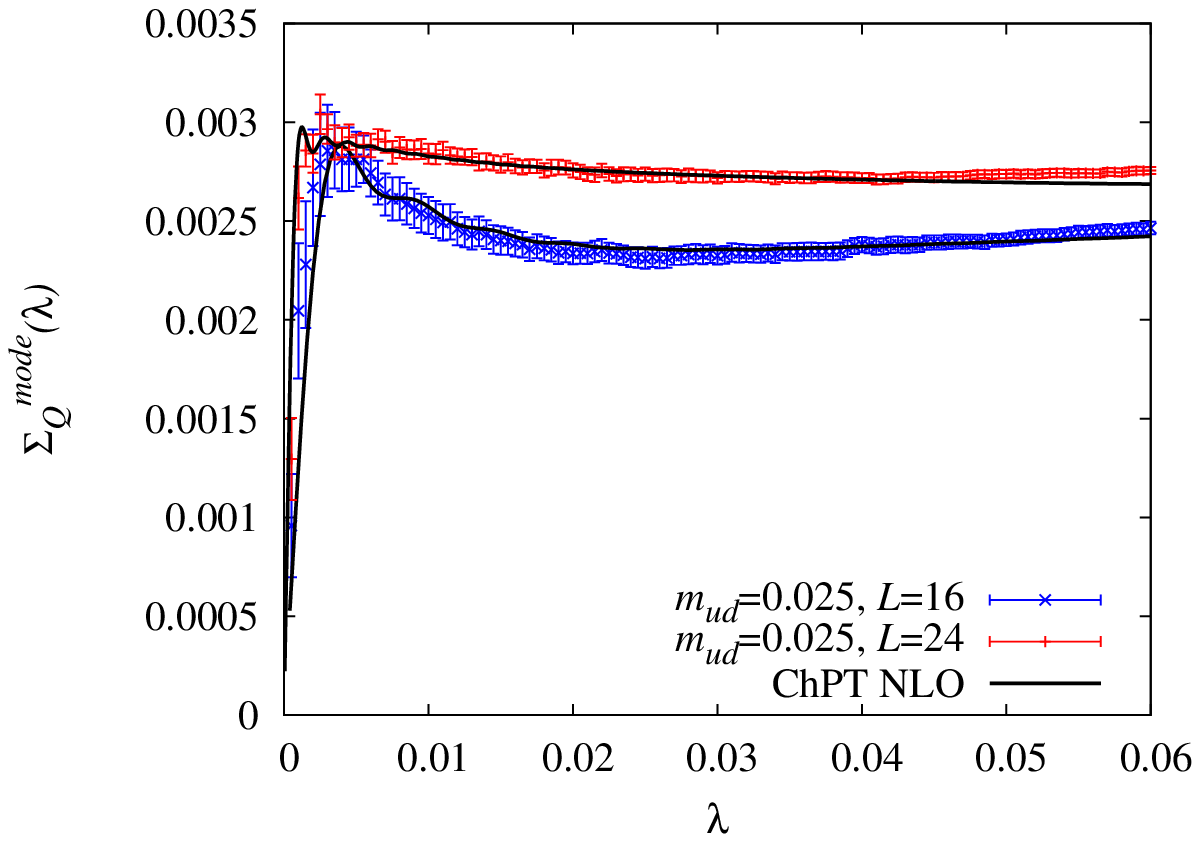}
  \caption{
    The spectral density at $m=0.025$ and $L=24$ (top)
    and comparison of $\Sigma_Q^{mode}(\lambda)$ at $L=24$ and 16 (bottom). 
    In the ChPT curves, the same values of $\Sigma_{eff}$ 
    (but the volume dependence is corrected within ChPT) and $F$
    are used as inputs.
  }
  \label{fig:rhoV}
\end{figure*}

One of the significant consequences of the ChPT formula (\ref{eq:rho})
is that the spectral function for different topological charge $Q$ and
volume $V$ should be described by the same set of the parameters, 
{\it i.e.} $\Sigma_{\mathrm{eff}}$ and $F$.
This provides a highly non-trivial cross-check of the formula.
For this purpose we produced data 
at non-zero topological charge $Q=1$.
The results are shown in Fig.~\ref{fig:rhoQ}.
Here the curves of the NLO ChPT is drawn with inputs from the $Q=0$
data and there is no further free parameter to adjust.
The good agreement below $\lambda\simeq$ 0.03 gives further
confidence on the analysis.

A similar check can be done with the lattice data obtained from a
larger volume lattice $24^3\times 48$, for which the data are shown in
Fig.~\ref{fig:rhoV}.
The comparison is a bit more tricky for different volumes, because the
definition of $\Sigma_{\mathrm{eff}}$ (\ref{eq:Sigmaeff}) depends on $V$.
Namely the function $\bar{\Delta}(M^2)$ contains $\bar{g}_1(M^2)$,
which represents the finite volume effect.
It is possible to convert the value of $\Sigma_{\mathrm{eff}}$ for
different volumes.
If we convert the result at $L=24$, $\Sigma_{\rm eff}=0.00306(7)$ to
the one on a $L=16$ lattice, it becomes 0.00341(18), which may be
compared with the independent calculation at $L=16$ at the same sea
quark mass $m=0.025$, which is 0.0333(18).
Therefore, the finite volume scaling is confirmed at least on two
different volumes, whose difference is a factor of 3.

The curves in Figures~\ref{fig:rho015}--\ref{fig:rhoV}
are drawn using the $N_f=2+1$ ChPT results, but we found 
the difference from $N_f=2$ ChPT formula is hardly
visible in the scale of this plot, which confirms decoupling
of the strange quark from the low energy theory.

From these analysis the values of $\Sigma_{\rm eff}$ and $F$ are
extracted for each sea quark mass.
Note that $\Sigma_{\rm eff}$ is extracted at the NLO accuracy, 
while the value of $F$, which first appears at the NLO term,
might have larger systematic corrections from NNLO contributions. 
We find that the results for $\Sigma_{\rm eff}$ are stable under 
change of two reference points in a range $\lambda < 0.03$.
As noted above, there is little difference between 
$N_f=2$ and $N_f=2+1$ formulae;
$\Sigma_{\rm eff}$ and $F$
are almost equal well within the statistical error.
The difference between $m_s=0.080$ and $m_s=0.100$
is even weaker.
In the following analysis, we concentrate on the data at $m_s=0.080$.


\subsection{Chiral extrapolation of $\Sigma_{\rm eff}$}
We next consider the sea quark mass dependence of $\Sigma_{\rm eff}$.
As shown in (\ref{eq:Sigmaeff}), $\Sigma_{\rm eff}$ is a function of 
$\Sigma$, $F$, $L_6$, which can be determined from the lattice data.
The chiral condensate $\Sigma$ thus obtained should have the NLO accuracy.
In the fitting of the lattice data,
we attempt
(A) 3-parameter ($\Sigma$, $F$, $L_6$) fit without any inputs
and (B) 2-parameter ($\Sigma$, $L_6$) fit with 
$F=0.0410$ (for $N_f=3$ ChPT) or with $F=0.0406$ (for $N_f=2$ ChPT).
These values of $F$ correspond to the chiral limit of $F$ extracted
from the analysis of the spectral function.

\begin{figure}[tbp]
  \centering
  \includegraphics[width=10cm]{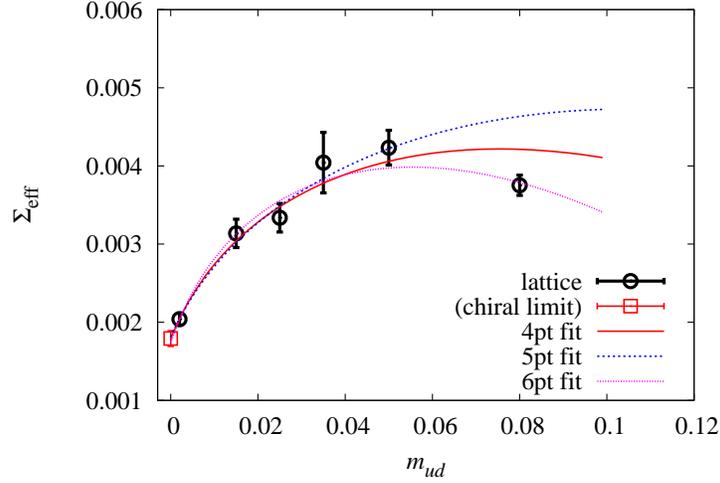}
  \caption{
    Three parameter fit of $\Sigma_{\rm eff}$.
    The $N_f=3$ ChPT formula is used.
  }
  \label{fig:SigmaefffitNf3}
\end{figure}

The fitting is shown in Fig.~\ref{fig:SigmaefffitNf3}
for the case (A) with the $N_f=3$ ChPT formula.
We use the lightest 4, 5, and 6 data points.
All the curves are consistent with the lattice data used in the fit
and in fact the $\chi^2$ per degrees of freedom is reasonable (between
0.6 and 1.5).
A remarkable fact is that the chiral limit (shown by a square) is not
sensitive to the number of data points used.
The chiral limit is very stable because of the presence of the
$\epsilon$-regime data point.
Similar curves are obtained for the case with $N_f=2$ and for the case
(B). 
With the 2-parameter fit (the case(B)) the heaviest data point cannot
be well described, {\it i.e.} $\chi^2/\mathrm{d.o.f.}$ is about 2.5.

From these curves, one can extract the low energy constants of ChPT.
Note in the case of $N_f=3$ ChPT, we have two limits of chiral
condensate: $\Sigma^{N_f=3}$, where ``three'' flavor massless limit is
taken, and $\Sigma^{\rm phys}$, which is a two-flavor chiral limit
with strange quark mass fixed at a finite value $m_s=0.08$. 
As already mentioned, the strange quark dependence is so small
that the difference from the value at the physical 
strange quark mass is negligible.
The extracted values of $\Sigma^{\rm phys}$ 
are stable against the different choice of fitting function and
fitting range, while $\Sigma^{N_f=3}$ shows strong sensitivity to
them. 
It means that the determination of $\Sigma^{N_f=3}$ is not feasible
with our current data set.
This is natural because the strange quark mass dependence is not
well controlled by the lattice data.
On the other hand, the determination of $\Sigma^{\rm phys}$ is very
stable, thanks to the $\epsilon$-regime data point.
Our estimate of systematic effects due to the chiral extrapolation
is $\sim 2$ \%.

From the above analysis, we determine the low-energy constants for
2+1-flavor QCD as
\begin{eqnarray}
\Sigma^{\rm phys} &=& 0.00186(10)(44) \sim [226(4)(18) \mbox{MeV}]^3,\\
F &=& 0.0406(05)(41) \sim 74(1)(8)\mbox{MeV},\\
L^r_6(770~{\rm MeV}) &=& -0.00011(25)(11),
\end{eqnarray}
where the first error is statistical and 
the second error is systematic, respectively.

To obtain the final result, 
we convert the value of $\Sigma^{\rm phys}$
to the definition in the $\overline{\mathrm{MS}}$ scheme, 
by using the non-perturbative renormalization factor 
\cite{Noaki:2009xi}calculated 
through the RI/MOM scheme \cite{Martinelli:1994ty}.
The result \cite{:2009fh}, $\Sigma^{\rm phys}$ 
in the limit of $m_{ud}=0$ and $m_s$ fixed at
its physical value,  is 
\begin{equation}
  \label{eq:final}
  \Sigma^{\overline{\mathrm{MS}}}(\mathrm{2~GeV}) = 
  [242(04)(^{+19}_{-18})~\mbox{MeV}]^3.
\end{equation}

Let us here discuss possible systematic errors in (\ref{eq:final}).
Since our lattice studies are done at only one value of $\beta$,
it is difficult to estimate the discretization errors.
But it should be partly reflected 
in the mismatch of the observables measured in different ways.
We here estimate it from a mismatch of the lattice spacing;
0.1003(46) fm from the pion decay constant \cite{Noaki} and
0.1087(15) fm from the $\Omega$ baryon mass \cite{Noaki2}.
This 7.4\% deviation is added in the systematic error.
The systematic error due to finite volume is estimated as $\sim 1.4$\%
using the lattice data at two different volumes.

%% file: sec5.tex
\section{Summary and Conclusion}
\label{sec:summary}

In this talk, a study of the spontaneous chiral symmetry breaking
performed by the JLQCD and TWQCD collaborations has been presented.
We discussed that fixing topology is an essential part 
for the dynamical overlap fermion simulations on the lattice.
By reducing the numerical cost with the topology fixing determinant,
we have performed the first large-scale simulations of 
dynamical overlap quarks.
The up and down quark masses are reduced close to the physical point.
We have then discussed that the global topology, 
as well as the finite volume effect,
can be described well within the chiral perturbation theory.
In fact, we have found a good agreement of our lattice data 
for the Dirac operator spectrum with the ChPT predictions even in the
region where its finite size effect is large.
We extract the chiral condensate in 2+1-flavor QCD.

\begin{acknowledgments}
  The author thanks P.H.~Damgaard and members of JLQCD and TWQCD collaborations
  for useful discussions.
  Numerical simulations are performed on IBM System Blue Gene
  Solution at High Energy Accelerator Research Organization
  (KEK) under a support of its Large Scale Simulation
  Program (No. 06-13).
  This work was supported by the Global COE program
  of Nagoya Univ. ``Quest for Fundamental Principles in the Universe''
  from JSPS and MEXT of Japan.
\end{acknowledgments}